\documentclass[twocolumn]{aastex6}

\shorttitle{Discovery of a Compact Companion}
\shortauthors{Stephen R. Kane et al.}

\begin{document}

\title{Discovery of a Compact Companion to a Nearby Star}

\author{
  Stephen R. Kane\altaffilmark{1},
  Paul A. Dalba\altaffilmark{1},
  Jonathan Horner\altaffilmark{2},
  Zhexing Li\altaffilmark{1},
  Robert A. Wittenmyer\altaffilmark{2},
  Elliott P. Horch\altaffilmark{4},
  Steve B. Howell\altaffilmark{3},
  Mark E. Everett\altaffilmark{5}
}
\altaffiltext{1}{Department of Earth and Planetary Sciences,
  University of California, Riverside, CA 92521, USA}
\altaffiltext{2}{Centre for Astrophysics, University of Southern
  Queensland, Toowoomba, QLD 4350, Australia}
\altaffiltext{3}{NASA Ames Research Center, Moffett Field, CA 94035,
  USA}
\altaffiltext{4}{Department of Physics, Southern Connecticut State
  University, New Haven, CT 06515, USA}
\altaffiltext{5}{National Optical Astronomy Observatory, 950 N. Cherry
  Ave, Tucson, AZ 85719, USA}
\email{skane@ucr.edu}


\begin{abstract}

Radial velocity (RV) searches for exoplanets have surveyed many of the
nearest and brightest stars for long-term velocity variations
indicative of a companion body. Such surveys often detect
high-amplitude velocity signatures of objects that lie outside the
planetary mass regime, most commonly those of a low-mass star. Such
stellar companions are frequently discarded as false-alarms to the
main science goals of the survey, but high-resolution imaging
techniques can be employed to either directly detect or place
significant constraints on the nature of the companion object. Here,
we present the discovery of a compact companion to the nearby star
HD~118475. Our Anglo-Australian Telescope (AAT) RV data allow the
extraction of the full Keplerian orbit of the companion, found to have
a minimum mass of 0.445~$M_\odot$. Follow-up speckle imaging
observations at the predicted time of maximum angular separation rule
out a main-sequence star as the source of the RV signature at the
3.3$\sigma$ significance level, implying that the companion must be a
low-luminosity compact object, most likely a white dwarf. We provide
an isochrone analysis combined with our data that constrain the
possible inclinations of the binary orbit. We discuss the eccentric
orbit of the companion in the context of tidal circularization
timescales and show that non-circular orbit was likely inherited from
the progenitor. Finally, we emphasize the need for utilizing such an
observation method to further understand the demographics of white
dwarf companions around nearby stars.

\end{abstract}

\keywords{white dwarfs -- techniques: radial velocities -- techniques:
  high angular resolution -- stars: individual (HD~118475)}


\section{Introduction}
\label{intro}

Over the last decade, large-scale exoplanet surveys have become
increasingly common. Whilst transit surveys, such as {\it Kepler}
\citep{borucki2010} and {\it TESS} \citep{ricker2015} are focused on
the search for small planets orbiting close to their host stars, a
number of long-term radial velocity (RV) surveys continue to scour the
sky, achieving particular sensitivity to long-period giant planets
\citep{bonfils2013a,butler2017}, analogous to the solar system's
Jupiter and Saturn \citep{wittenmyer2016c}. Since RV semi-amplitudes
decrease toward longer periods, the sensitivity of RV surveys likewise
shifts towards high masses with increasing orbital period
\citep{kane2007a}. Even though this limits the use of such surveys in
exploring the low-mass, distant planet regime, they remain ideally
suited as probes of the occurrence of objects that bridge the gap
between the planetary and stellar mass regimes. Such surveys are
therefore ideal for the study of the demographics in that region
\citep{duquennoy1991b,raghavan2010}. Combining RVs with high angular
resolution observations provides an additional avenue through which we
can test models of the frequency of companion mass for a variety of
stars \citep{kane2014c,crepp2016,wittrock2017}, including stars known
to host exoplanets \citep{kane2015c,roberts2015a,wittrock2016}.

The combination of imaging and RV techniques can further be used to
detect more exotic companions, such as compact objects. An example of
this is the detection of a white dwarf that was first identified
through the observation of a linear trend in the RVs measured for
HD~169889, before being directly observed using high-contrast imaging
\citep{crepp2018}. Numerous RV observations of white dwarfs have been
carried out \citep{barnbaum1992,maxted2000,rebassa2017}, but it is
relatively rare for white dwarfs to be identified using the RV method
due to the ambiguity of the orbital inclination.

The Anglo-Australian Planet Search (AAPS) is one of the longest
running RV searches for exoplanets, with a temporal baseline of 18
years \citep{wittenmyer2014a}. The results from this survey have
revealed numerous high-amplitude RV signatures that fall outside of
the planetary mass regime. Here, we present RV data for the star
HD~118475 that reveal a companion moving on a well constrained 2070
day period orbit. The minimum mass of the companion, 0.445~$M_\odot$,
would place it firmly in the stellar mass regime - comparable to the
mass of an early M dwarf. As the primary star is relatively nearby
($\sim$~33 pcs), the orbital separation of the secondary
($\sim$3.69~AU) corresponds to an angular separation of $0.11\arcsec$
at quadrature. Despite the fact that such a companion should be
readily detected, follow-up observations carried out close to the
predicted maximum angular separation using the Differential Speckle
Survey Instrument (DSSI) on Gemini-South rule out such a main-sequence
star as the companion. As a result, we conclude that the companion
must instead be a compact object, such as a white dwarf.

The structure of this paper is laid out as follows. In
Section~\ref{rv} we provide the RV data along with the best-fit
Keplerian orbital solution. Section~\ref{imaging} includes the details
of the DSSI observations and the reduced data confirming the exclusion
of a bright companion. In Section~\ref{wd}, the significance of a
null detection is quantified in the context of stellar isochrones, and
we use the compact nature of the companion to place additional
constraints on the orbital inclination of the system. We provide
concluding remarks in Section~\ref{conclusions} along with suggestions
for further observations.


\section{Companion Mass and Orbit}
\label{rv}

The RV observations of HD~118475 were acquired using the UCLES
high-resolution spectrograph \citep{diego1990} on the 3.9m
Anglo-Australian Telescope (AAT). The observations and data reduction
are able to routinely achieve a velocity precision of 2--3~m\,s$^{-1}$
through the use of an iodine absorption cell
\citep{valenti1995b,butler1996} that provides wavelength calibration
from 5000 to 6200\,\AA. These data have been successfully used over
many years to detect planetary-mass companions to nearby stars
\citep{butler2006,wittenmyer2017c}. The AAT dataset for HD~118475
consists of 11 measurements acquired over a time baseline of $\sim$12
years. These data are shown in Table~\ref{vels}.

\begin{figure}
  \begin{center}
    \includegraphics[width=8.5cm]{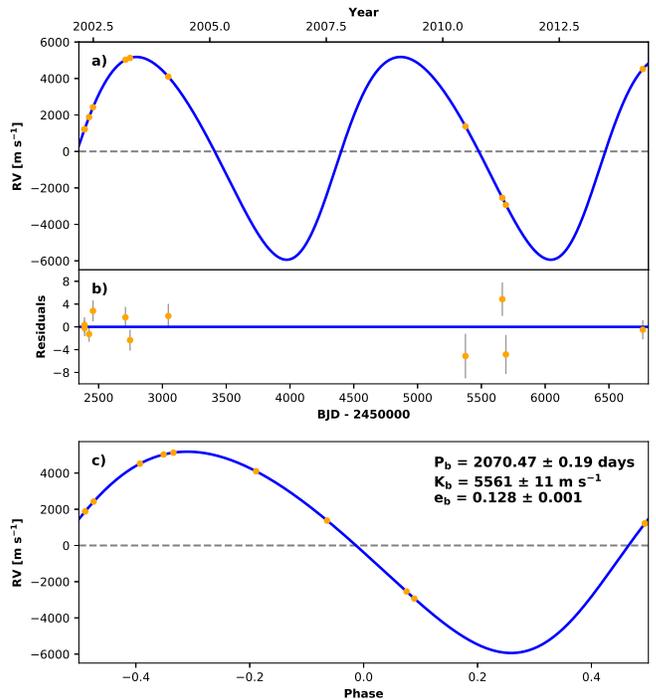}
  \end{center}
  \caption{AAT RV data for HD~118475 (shown as yellow points) and the
    best-fit model (shown as a blue solid curve). The unphased data
    with residuals are shown in the top panel, and the phased data are
    shown in the bottom panel.}
  \label{rvs}
\end{figure}

To fit the RV data, we used a modified version of the RadVel package
developed by \citet{fulton2018a}. The RadVel code was originally
designed for only planetary mass companions because of approximations
related to the mass and semi-major axis calculations. Our modification
of the RadVel code allows for more massive companions by including the
mass of the secondary in the equations that extract the companion
minimum mass and semi-major axis from the Keplerian orbital
parameters. This modification to RadVel allowed us to calculate the
correct mass and orbital semi-major axis of the companion. Shown in
Figure~\ref{rvs} are the unphased data with residuals (top panel) and
the phased data (bottom panel) along with the best-fit model (solid
curve). An offset of $1883.0\pm5.6$~m\,s$^{-1}$ was applied to the
data during the fit to set the zero-point to the mean value of the
model. The extracted companion parameters from the RV fit are shown in
Table~\ref{params}, where $T_c$ is the time of inferior
conjunction. Also shown in Table~\ref{params} are the host star
properties from the Spectroscopic Properties of Cool Stars (SPOCS)
catalog \citep{valenti2005} and the derived companion properties of
minimum mass and semi-major axis. Note that the uncertainties on the
mass of the secondary are primarily correlated with the mass
uncertainties of the host star.

\begin{deluxetable}{ccc}
  \tablewidth{0pc}
  \tablecaption{\label{vels}HD 118475 AAT Radial Velocities}
  \tablehead{
    \colhead{Date} &
    \colhead{RV} &
    \colhead{$\sigma$} \\
    \colhead{(BJD -- 2450000)} &
    \colhead{(m\,s$^{-1}$)} &
    \colhead{(m\,s$^{-1}$)}
  }
  \startdata
  2389.12939 &  -675.97 &     1.36 \\
  2390.06030 &  -658.42 &     1.41 \\
  2425.04517 &    -1.36 &     1.34 \\
  2455.95264 &   542.14 &     1.84 \\
  2710.18823 &  3142.47 &     1.84 \\
  2746.11157 &  3243.84 &     1.85 \\
  3046.20020 &  2219.17 &     2.13 \\
  5374.98975 &  -506.48 &     3.92 \\
  5664.16113 & -4427.03 &     2.93 \\
  5692.07959 & -4815.68 &     3.42 \\
  6765.16699 &  2636.41 &     1.69 \\
  \enddata
\end{deluxetable}

\begin{deluxetable}{lr}
  \tablecolumns{2}
  \tablewidth{0pc}
  \tablecaption{\label{params}HD~118475 System Properties}
  \tablehead{
    \colhead{Parameter} & 
    \colhead{Value}
  }
  \startdata
  \sidehead{\bf Host Star}
  ~~~~$V$                         & 6.97 \\
  ~~~~$d$ (pcs)                   & 32.9 \\
  ~~~~$M_\star$ ($M_\odot$)       & $1.12 \pm 0.11$ \\
  ~~~~$T_\mathrm{eff}$ (K)        & $5898 \pm 44$ \\
  ~~~~$\log g$                    & $4.36 \pm 0.06$ \\
  ~~~~{[Fe/H]}                    & $0.07 \pm 0.03$ \\
  \hline
  \sidehead{\bf Companion Measured}
  ~~~~$P$ (days)                  & $2070.47^{+0.19}_{-0.2}$ \\
  ~~~~$T_c$ (BJD)                 & $2455507.9^{+1.0}_{-1.1}$ \\
  ~~~~$e$                         & $0.128 \pm 0.001$ \\
  ~~~~$\omega$ (deg)              & $237.7 \pm 0.3$ \\
  ~~~~$K$ (m\,s$^{-1}$)           & $5561 \pm 11$ \\
  \hline
  \sidehead{\bf Companion Derived}
  ~~~~$M_B \sin i$ ($M_\odot$)    & $0.445 \pm 0.025$ \\
  ~~~~$a$ (AU)                    & $3.69 \pm 0.11$ \\
  \hline
  \sidehead{\bf Measurements and Model}
  ~~~~$N_{\mathrm{obs}}$          & 11 \\
  ~~~~rms (m\,s$^{-1}$)           & 2.93 \\
  ~~~~$\chi^2_{\mathrm{red}}$     & 3.27
  \enddata
\end{deluxetable}


\section{Direct Imaging Observations}
\label{imaging}

\begin{figure*}
  \begin{center}
    \begin{tabular}{cc}
      \includegraphics[width=7.0cm]{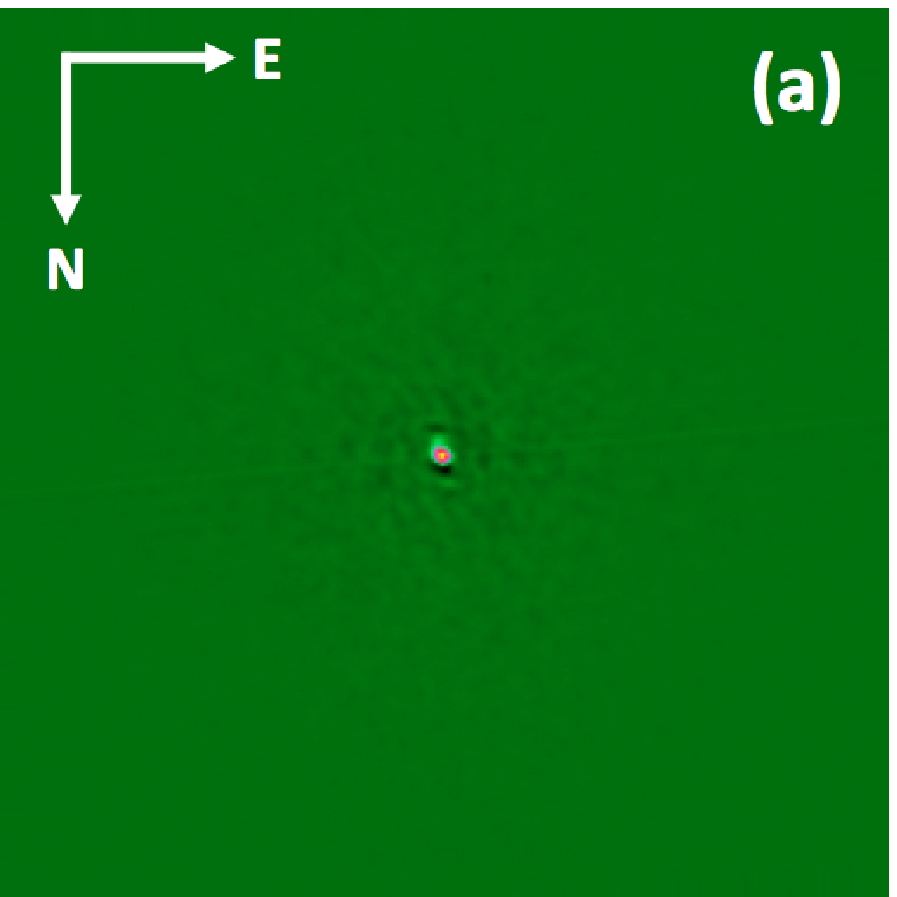} &
      \includegraphics[width=10.0cm]{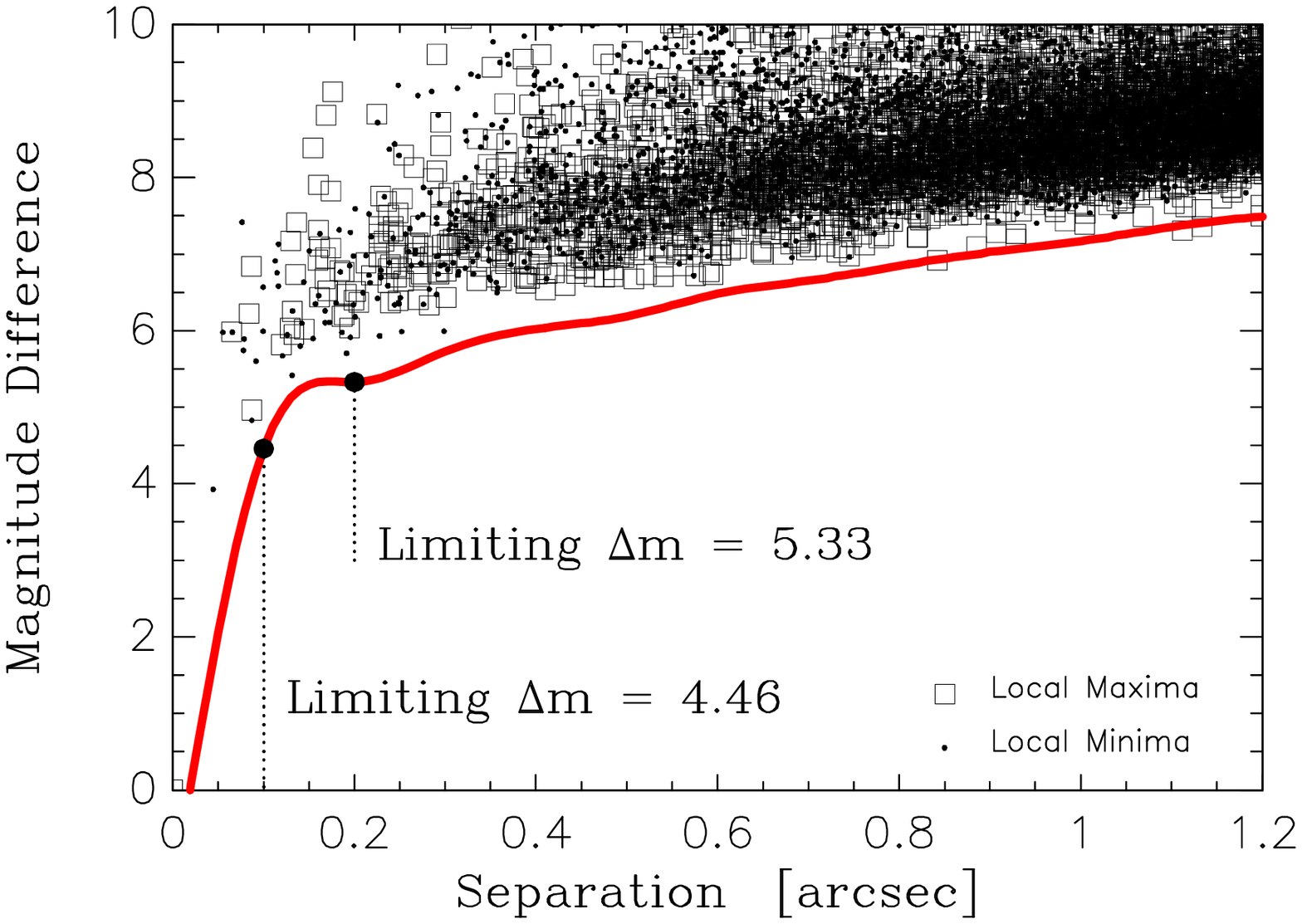} \\
      \includegraphics[width=7.0cm]{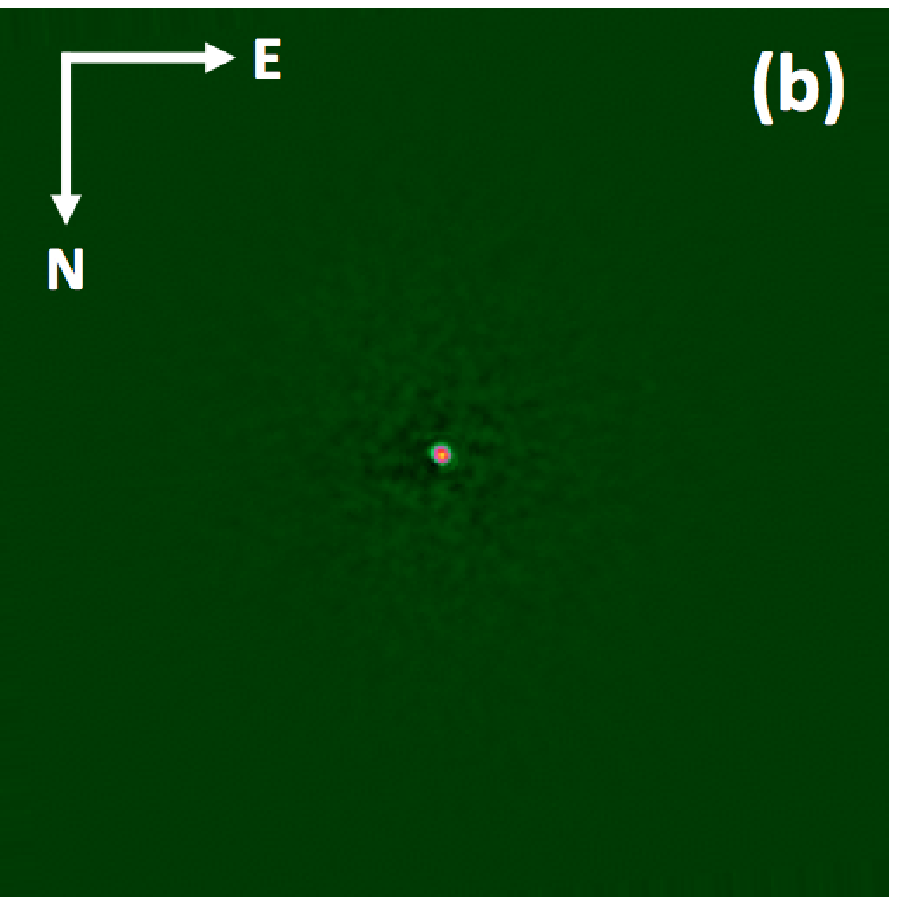} &
      \includegraphics[width=10.0cm]{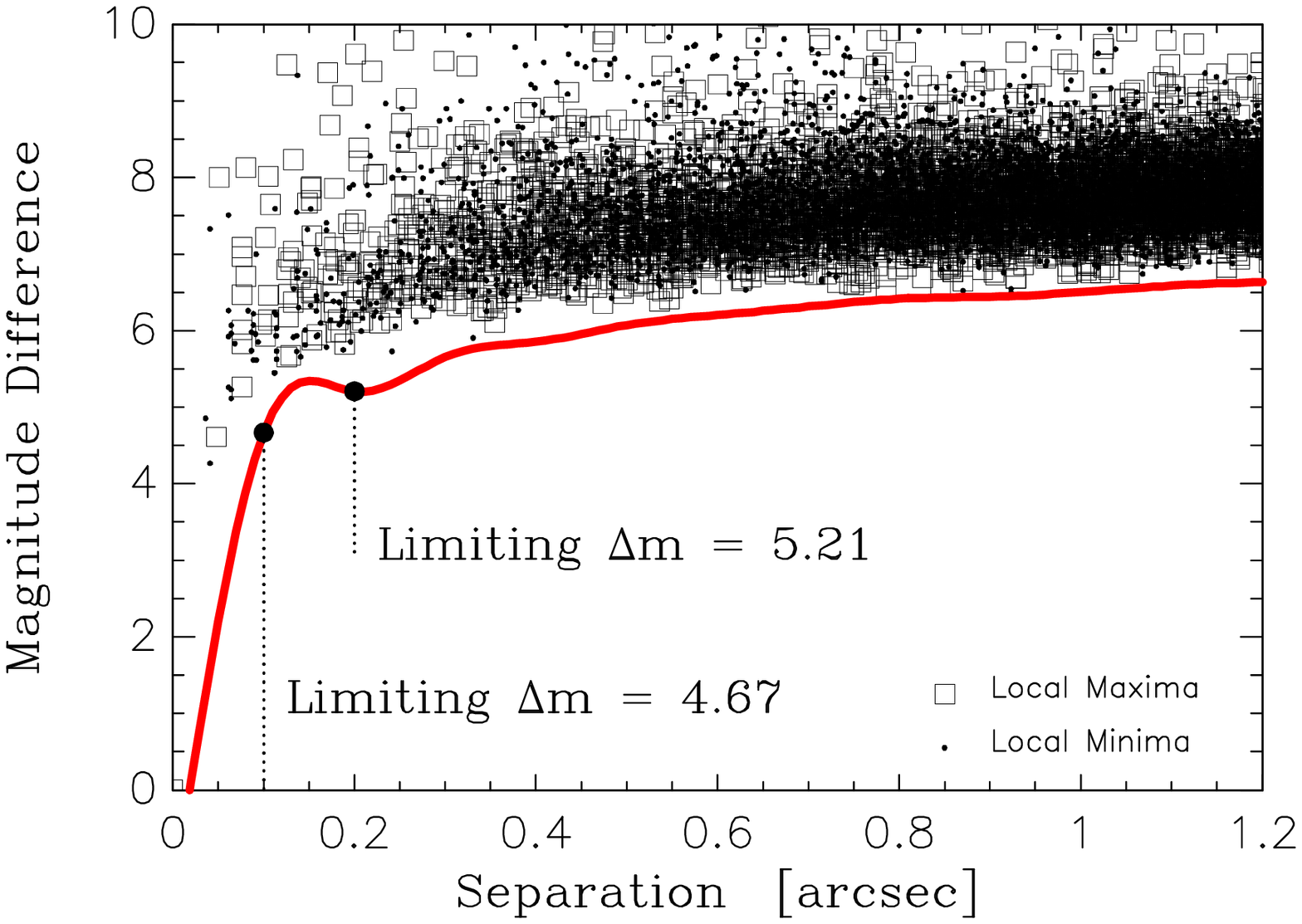}
    \end{tabular}
  \end{center}
  \caption{The DSSI images (left) and detection limit plots (right)
    for the 692~nm (top) and 880~nm (bottom) bandpasses
    respectively. The images on the left are marked as (a) and (b) for
    692~nnm and 880~nm respectively. The field-of-view for the camera
    is 2.8$\times$2.8$\arcsec$, north is down, and east is to the
    right. The limiting magnitude data shown in the right panels
    include cubic spline interpolations (red solid line) of the
    5$\sigma$ detection limit.}
  \label{dssi}
\end{figure*}

DSSI is a dual-channel speckle imaging system in which each channel
records speckle patterns in narrowband filters with central
wavelengths of 692 and 880~nm. The instrument is described in more
detail by \citet{horch2009}, including specifics regarding data
reduction and performance statistics. HD~118473 was observed using
DSSI on Gemini-South during the night of June 7, 2017 (BJD =
2457911). Shown in the left panels of Figure~\ref{dssi} are the DSSI
images using the 692~nm (top) and 880~nm (bottom) filters, where the
contrast of the images have been fixed to maximize the dynamic range
of flux due to possible companions. The limiting magnitude curves are
shown in the right panels of Figure~\ref{dssi} for the 692~nm (top)
and 880~nm (bottom) filters. These limiting magnitude figures plot the
magnitude difference between local maxima and minima as a function of
the separation from the primary star and include a cubic spline
interpolation (red solid line) of the 5$\sigma$ detection limit for
the full range of angular separations represented.

According to the orbital solution shown in Table~\ref{params}, the
DSSI observation occurred $\sim$333 days past the passage of inferior
conjunction. Using the formalism developed by \citet{kane2013c} and
\citet{kane2018c}, we calculate the angular separation between the
primary star and the companion at the time of DSSI observation to be
$\sim$0.09$\arcsec$, compared with the maximum angular separation for
the companion of 0.11$\arcsec$. The resulting detection limits of the
companion at the time of observation are discussed in detail in the
following section.


\section{The White Dwarf Hypothesis}
\label{wd}

\begin{figure*}
  \begin{center}
    \begin{tabular}{cc}
      \includegraphics[width=8.5cm]{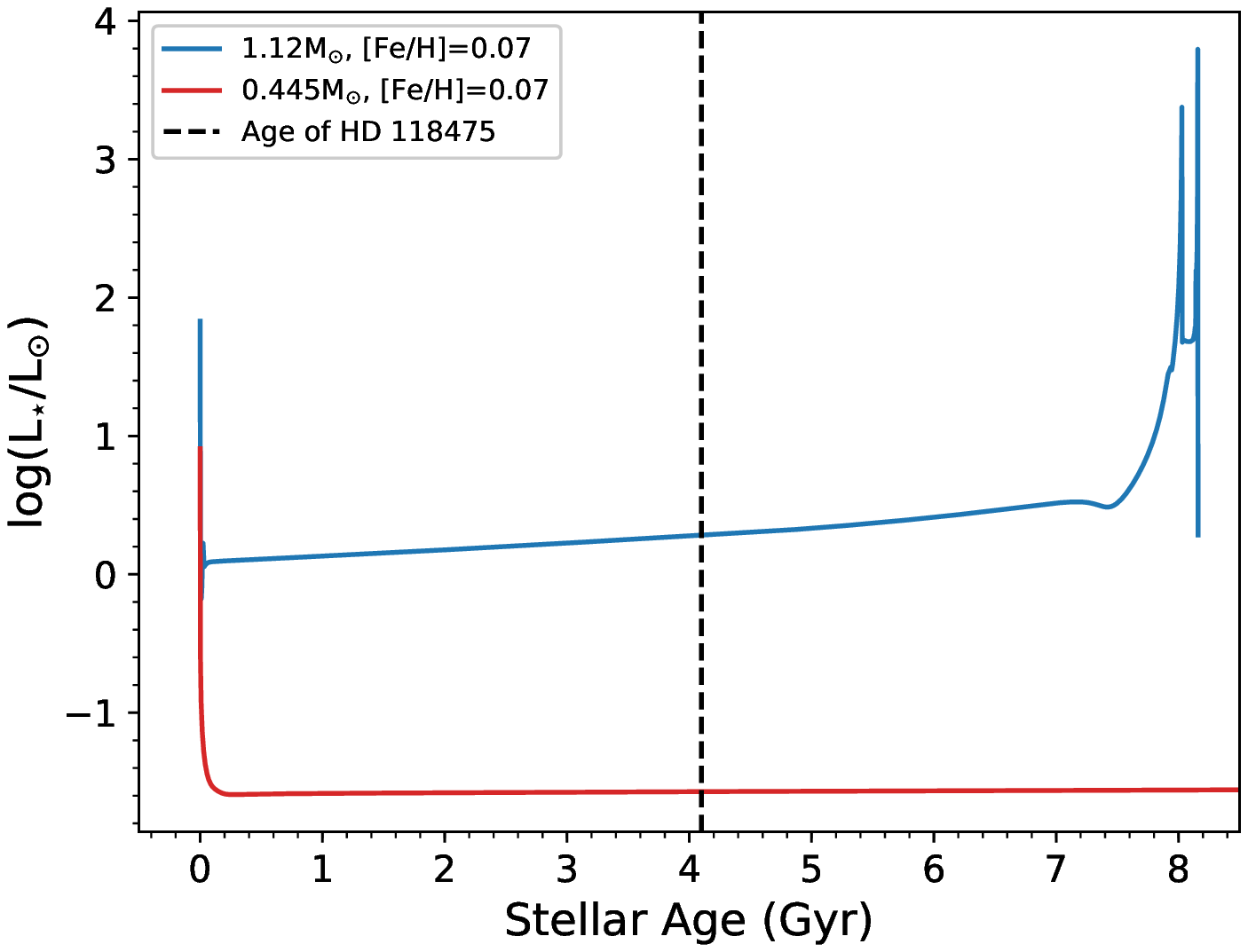} &
      \includegraphics[width=8.5cm]{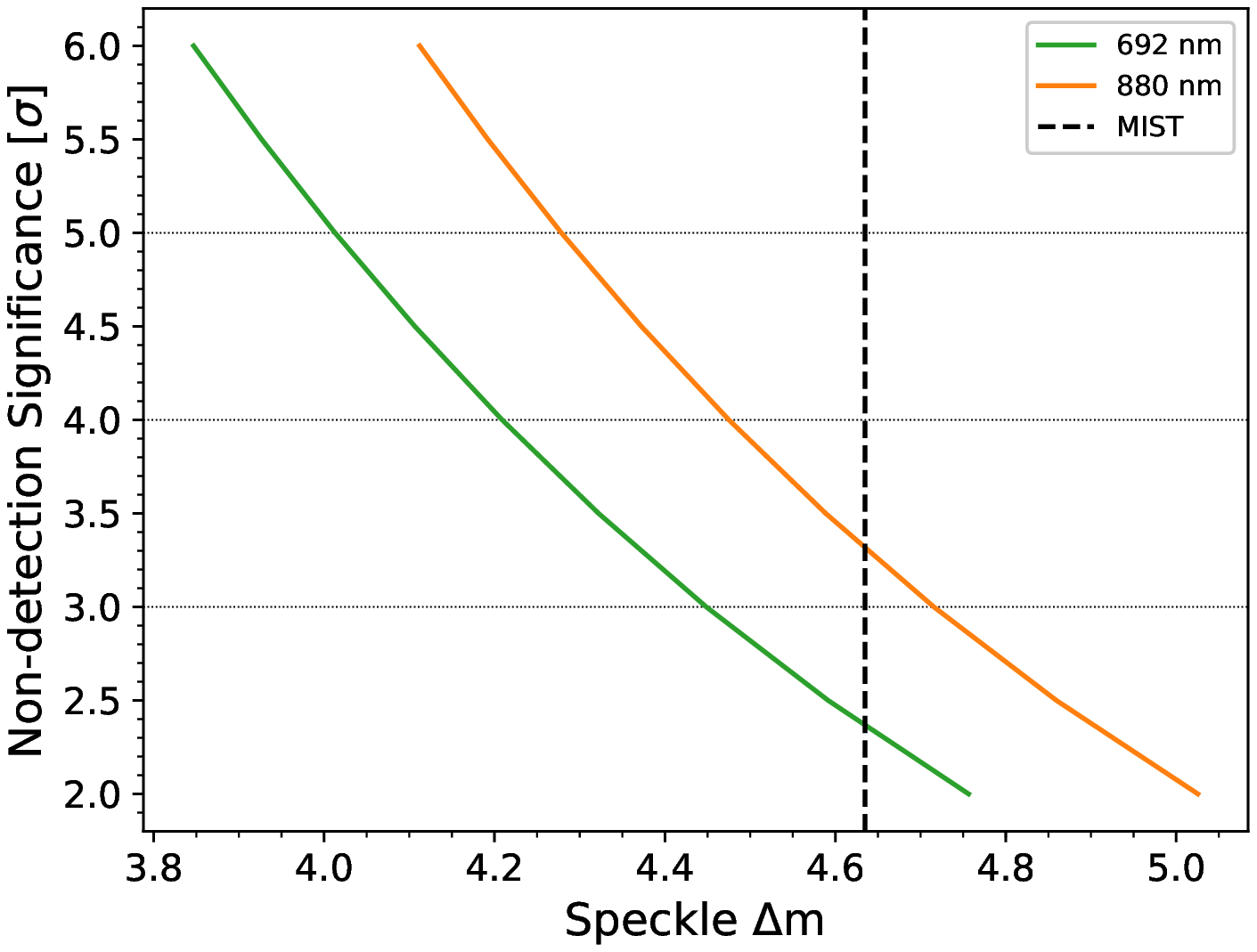}
    \end{tabular}
  \end{center}
  \caption{Left: evolutionary tracks for the primary and the
    secondary---assuming the secondary is an M dwarf---from MIST. The
    system age \citep[$\sim$4.1 Gyr,][]{valenti2005} is indicated by
    the dashed line. At this age, the M dwarf secondary would be 4.63
    magnitudes fainter than the primary. Right: the non-detection
    significance inferred from the 692~nm and 880~nm observations. The
    dashed line is drawn at 4.63 magnitudes, which is ruled out to a
    significance of 2.37$\sigma$ and 3.32$\sigma$ for 692 and 880 nm,
    respectively.}
  \label{isochrones}
\end{figure*}

The imaging observations detailed in Section~\ref{imaging} should be
more than capable of resolving a main-sequence companion of the mass
required to explain the RV observations described in
Section~\ref{rv}. We demonstrate this by modeling the luminosity
evolution of the primary and a 0.445~$M_\odot$ (M dwarf) secondary
using the MESA Isochrones and Stellar Tracks (MIST) models
\citep{choi2016,dotter2016}. As shown in the left panel of
Figure~\ref{isochrones}, at the current age of the system
\citep[$\sim$4.1 Gyr,][]{valenti2005}, an M dwarf companion would be
71.4 times less luminous than the primary (corresponding to a
magnitude difference of 4.63 magnitudes). With this result, we can
confidently rule out the hypothesis that the companion is an early M
dwarf to the 2.37$\sigma$ level in the 692~nm observation, and to the
3.32$\sigma$ level in the 880 nm observation (see the right panel of
Figure~\ref{isochrones}), for the angular separation at the time of
observation (see Section~\ref{imaging}).

At first glance, the uncertainty in the mass of the companion might
appear to offer a ready explanation for the non-detection. Since the
luminosity of a given star is, to first order, proportional to its
mass to the fourth power \citep{smith1983}, one might think that a
relatively small reduction in the mass of the secondary could be
sufficient to lower its luminosity to a level where it would not be
detectable. However, the uncertainty in the mass of the secondary is
strongly correlated with the uncertainty in the mass of the
primary. In fact, the derived uncertainty in the secondary's mass is
smaller, as a fraction of the total mass, than the uncertainty on the
primary. Therefore, the only way to reduce the companion's mass would
be to reduce the primary's mass. In doing so, the contrast ratio
between them would still be such that the companion would be readily
detectable in imaging observations at 880 nm.

Interestingly, the fact that the companion must be a compact object in
turn means that the system cannot be edge-on to our line of
sight. Given the age of the system ($\sim$4.1~Gyr), we can place
limits on the minimum mass that such a compact companion could have,
on the basis that such a body must have passed through the entirety of
its main-sequence evolution. We modeled the evolution of stars of
varying mass, assuming that the initial metallicity of the companion
matched that of the primary ([Fe/H]$=$0.07), using MIST. As a function
of initial progenitor mass, we noted the age at which the post-main
sequence mass-loss rate declined to approximately zero. This age was
coincident with the beginning of the white dwarf cooling phase. In
this manner, we determined that the lowest-mass progenitor for a white
dwarf companion that would have completed its evolution within the
4.1-Gyr age of the system would be a 1.38~$M_\odot$ star. According to
the MIST tracks shown in Figure~\ref{isochrones}, such a star would
leave a white dwarf of mass 0.559~$M_\odot$. Comparing this mass with
the $M_B \sin i$ of 0.445~$M_\odot$ allows us to estimate the range of
orbital inclinations that are allowed for our solution. We find that
orbital inclinations between 90$\degr$ (edge-on) and 52.8$\degr$ are
excluded by the compact nature of the white dwarf. Thus, the system is
not old enough to have produced such a low-mass compact
companion. There is obviously a chance that the unseen companion could
be more massive still. A white dwarf can be no more massive than the
Chandrasekhar limit of 1.4~$M_\odot$. If the companion is truly a
white dwarf, this means that the inclination cannot be less than
18.5$\degr$, as such an inclination would require the mass to exceed
the Chandrasekhar limit.

This opens up the interesting, but unlikely possibility that the
unseen companion is either a neutron star or black hole. Our
observations do not rule out such an eventuality. We note that the
theoretical maximum mass for a neutron star is of order
$\sim$3~$M_\odot$ which corresponds to an orbital inclination for the
system of $\sim$8.5$\degr$, beyond which the unseen companion must be
a black hole.

A fascinating aspect of the system is the significantly non-zero
eccentricity of the compact companion. Given the age of the system and
the reasonable assumption that the progenitor of the compact object
had a higher initial mass, conservation of angular momentum would
require that the progenitor had a smaller separation from the current
primary than is currently observed. In that case, one may presume that
tidal circularization would have produced a circular orbit before the
progenitor departed from the main-sequence. To estimate the range of
semi-major axes over which we can expect circularization to have
occurred, we used the expression for the turbulent dissipation
circularization timescale provided in Equation~(4.13) by
\citet{zahn1977}. For simplicity, we assumed a mass ratio of unity, a
primary radius of 1~$R_\odot$, primary mass of 1.12~$M_\odot$ (see
Table~\ref{params}), primary luminosity of 1~$L_\odot$, and an apsidal
motion constant for the second harmonic of $k_2 = 0.01444$ for a
polytropic index of $n = 3$ \citep{brooker1955}. Based on these
values, we find that the main-sequence progenitor of the compact
companion would not have become tidally circularized before leaving
the main-sequence unless it was located within $\sim$0.1~AU of the
current primary star, which is unlikely given its present semi-major
axis of 3.69~AU. However, departure from the main-sequence into the
red giant phase can have a dramatic effect on the tidal
circularization timescale for binary systems. Using the methodology of
\citet{verbunt1995} and our estimated minimum mass of the progenitor
(1.38~$M_\odot$), we estimate that tidal circularization of the
current primary by the white dwarf progenitor would have occurred out
to orbital periods of $\sim$1550~days, or 3.55~AU. However,
outspiralling of the progenitor due to mass loss during the transition
from the asymptotic giant branch to a white dwarf means that the
stellar components were originally much closer together. Given that
the current semi-major axis of the system is 3.69~AU (see
Table~\ref{params}) and the original separation of the system would
have been much smaller, tidal circularization should certainly have
occurred. Thus, the eccentricity of the companion is unlikely to have
been inherited from its orbit whilst on the main-sequence. The current
orbit may have been perturbed by a close stellar encounter or
additional companion in a wide orbit.

A particular issue that is raised by the discovery of this compact
companion is the completeness of white dwarf surveys in the solar
neighborhood. For example, it was found by \citet{tremblay2014} that
the results of various white dwarf surveys are consistent with each
other, but inconsistent with the expected population of white dwarfs
based upon the demographics and age of nearby stars. This suggests
that there remains a large fraction of stars for which their true
binarity remains unresolved. The observational methods used in this
work highlight an additional avenue through which the current dearth
of known white dwarfs around nearby stars may be mitigated.

In the coming years, data from the {\it Gaia} spacecraft
\citep{prusti2016} should provide a measurement of the amplitude of
the astrometric wobble imposed upon HD~118475 by its unseen
companion. Such data will provide an unequivocal answer to the true
mass of the companion. The RV observations made by the AAPS constrain
the line-of-sight motion of the star, whilst the astrometric data
obtained by {\it Gaia} will detail the motion at right angles to our
line of sight. By combining the two, the true system inclination will
be determined, which in turn will precisely constrain the companion
mass.


\section{Conclusions}
\label{conclusions}

Using RV observations of the star HD~118475 that span a period of 11
years, we find evidence of a massive ($M_B \sin i = 0.445$~$M_\odot$)
companion moving on a $\sim$2070 day orbit, corresponding to an
orbital semi-major axis of 3.69~AU. Typically, one would assume that
such a companion is most likely to be a previously undetected main
sequence star, with the calculated mass suggesting an early M
dwarf. We therefore carried out observations of the system using DSSI
on Gemini-South. With those observations, we can rule out a main
sequence companion to HD~118475 at the $\sim$3.3$\sigma$ level.

The source of the periodic RV signal observed for HD~118475 must
therefore be a compact object, most likely a white dwarf based on the
range of possible orbital inclinations. By considering the age of the
system ($\sim$4.1 Gyr), we determine that the minimum mass that such a
compact companion could have is $\sim$0.56~$M_\odot$. Compact
companions with a lower mass can be excluded on the basis that the
progenitor required for such a body would not have had time to evolve
beyond its main-sequence lifetime. The fact that the companion must be
more massive than $\sim$0.56 $M_\odot$ means that the system's orbital
plane cannot be edge-on to the line of sight. Indeed, orbital
inclinations greater than $\sim$53$\degr$ can be excluded on the basis
of the calculated minimum white dwarf mass.

We note that orbital inclinations between $\sim$8$\degr$ and
$\sim$18$\degr$ would suggest that the unseen companion is actually a
neutron star, while inclinations less than $\sim$8$\degr$ would
suggest a black hole. Such outcomes are unlikely, given the scarcity
of such massive compact objects, and so the white dwarf hypothesis
seems by far the most likely explanation for the non-detection of a
companion through direct imaging. In the future, the release of data
obtained by the {\it Gaia} mission will allow the orbital inclination
of the system to be determined with exquisite precision, which, in
combination with the existing RV data, will produce a measurement of
the companion's true mass. Furthermore, direct imaging experiments
with greater sensitivity capabilities are highly encouraged to attempt
to detect the unseen companion, and confirm its white dwarf nature. A
DA white dwarf has an absolute magnitude of $M_V \sim 12$, which
results in a required sensitivity of at least $\Delta m_V \sim 7$ for
a successful detection. More generally, our results serve as an
important reminder of the value of long-term RV exoplanet surveys, and
suggest that the data from such surveys should be revisited to examine
systems for which long-term, high-amplitude trends led to certain
targets being abandoned in favor of those more likely to yield
exoplanetary detections.


\section*{Acknowledgements}

This work is based on observations obtained at the Gemini-South
Observatory, which is operated by the Association of Universities for
Research in Astronomy, Inc., under a cooperative agreement with the
NSF on behalf of the Gemini partnership: the National Science
Foundation (United States), National Research Council (Canada),
CONICYT (Chile), Ministerio de Ciencia, Tecnolog\'{i}a e
Innovaci\'{o}n Productiva (Argentina), Minist\'{e}rio da Ci\^{e}ncia,
Tecnologia e Inova\c{c}\~{a}o (Brazil), and Korea Astronomy and Space
Science Institute (Republic of Korea). We acknowledge the traditional
owners of the land on which the AAT stands, the Gamilaraay people, and
pay our respects to elders past and present. The results reported
herein benefited from collaborations and/or information exchange
within NASA's Nexus for Exoplanet System Science (NExSS) research
coordination network sponsored by NASA's Science Mission Directorate.




\end{document}